\documentclass[manuscript,nonacm]{acmart}

\AtBeginDocument{%
  }

\renewcommand\footnotetextcopyrightpermission[1]{}

\usepackage{color}
\usepackage{graphicx}
\usepackage{subfigure} 
\usepackage{booktabs} 
\usepackage{multirow}
\usepackage{placeins}
\usepackage{float}

\begin{document}

\title{AI Humor Generation: Cognitive, Social and Creative Skills for Effective Humor}
\renewcommand{\shorttitle}{AI Humor Generation}

\author{Sean Kim}
\email{ssk2245@columbia.edu }
\affiliation{
    \institution{Columbia University}
    \city{New York}
    \state{New York}
    \country{USA}
}

\author{Lydia B. Chilton}
\email{chilton@cs.columbia.edu}
\affiliation{
    \institution{Columbia University}
    \city{New York}
    \state{New York}
    \country{USA}
}

 

\begin{abstract}
Humor is a social binding agent. It is an act of creativity that can provoke emotional reactions on a broad range of topics. Humor has long been thought to be “too human” for AI to generate. However, humans are complex, and humor requires our complex set of skills: cognitive reasoning, social understanding, a broad base of knowledge, creative thinking, and audience understanding. We explore whether giving AI such skills enables it to write humor. We target one audience: Gen Z humor fans. We ask people to rate meme caption humor from three sources: highly upvoted human captions, 2) basic LLMs, and 3) LLMs captions with humor skills. We find that users like LLMs captions with humor skills more than basic LLMs and almost on par with top-rated humor written by people. We discuss how giving AI human-like skills can help it generate communication that resonates with people. 

\end{abstract}




\maketitle

\section{Introduction}
Producing humor is a difficult yet quintessential endeavor that underpins the everyday human experience. Whether it is cracking a joke with coworkers, sending memes to friends, or playfully flirting with a romantic interest on a date, most people interface with humor every day in some capacity. We use humor to connect with people, to impress people, to point out absurdities, to make light of a bad situation, and to recognize universal struggles - big and small. 

Humor has long been thought to be “too difficult" or "too human” for AI to generate. Humor is complex, and there is no single formula for it. Different people find different things funny at different times. 
Humor seems to require social skills to know one's audience's point of view, understand human relationships, see things from multiple points of view, and judge what is socially appropriate to make fun of. 
Humor requires cognitive and reasoning skills because most jokes are clever in some way - they don't just say the obvious, they construct facts in ways that bring out absurdities, contradictions, and surprises.
Humor is a creative act, and it benefits from making keen observations, trying multiple angles, and multiple wordings.
Getting a person to do all these things is hard enough. Asking AI to do it is even harder.

However, generative AI has created new possibilities for computational humor. From training on large corpora of text, large language models have a broad base of knowledge and an implicit understanding of what humor is. However, even state-of-the-art models are known for being disappointing in their humor abilities~\cite{gptnotfunny}. Theories of humor indicate that there are patterns to humor~\cite{humor_Raskin2009}, which makes it theoretically possible for generative AI to implicitly learn these patterns, and its attempts at humor show that it generally understands the structure. And yet humor is complex enough that knowing the structure is not enough - it also requires human-like skills such as understanding people, culture, and the human experience. 
\color{black}

We test whether AI-generated humor that uses creative, social, and cognitive skills can come close to human performance. As a research context, we study the specific problem of generating humorous captions for images posted to Instagram. This is a popular humor type within Gen Z that is widely made and widely appreciated by a well-defined audience. There is ground truth (in the form of upvotes) for what captions the audience finds funny. It is a fairly complex form of humor that uses both visual and language skills to generate, it applies to a wide variety of input images and requires implicit Gen Z cultural knowledge to generate. 

Our method of generating humor uses three steps, following the divergent and convergent stages often used in cognitive models of creativity. 
First is an observation phase where AI takes in the image and makes careful observations of things in the photo.
Next is a divergent phase where AI generates multiple possible humorous angles. We have two strategies for this: one that poses humor strictly related to the image content and one that departs from the image content by 
bringing in relatable social conflicts that are analogous to it.
Next, a generation phase where based on the angles, we generate 30 possible captions. As with most ideation, we favor quantity and diversity over quality. 
Lastly, a ranking phrase where a ``Gen-Z humor expert'' agent rates the generated captions and selects the 5 best to return. 

We compare the HumorSkills generation method to two other sources of captions: 1) the five most upvoted comments on Instagram for the same image and 2) a state-of-the-art VLM (GPT-4o) with only prompt engineering. Through a humor rating survey of 20 images with 15 captions each, we found that HumorSkills is significantly funnier than the VLM baseline, and as funny as the top Instagram comments. HumorSkills captions were rated only 0.08 points lower on a 5-point scale with p=0.053, thus making HumorSkills not statistically less funny than the best human captions. 

We end with a discussion of how imbuing generative AI with multiple human-like skills can enhance its ability to do complex human communication. 
\section{Related Work}
\subsection{Humor Theory}

Humor is an intellectual challenge, humor has been studied by many great Western thinkers – Plato, Kant, Freud~\cite{sep_humor}.
Many theories have been proposed about why we have a sense of humor and what we find funny, and many cite social phenomena as the root cause.
Philosopher Daniel Dennett ~\cite{hurley2011inside} theorizes that the reason humans evolved humor was so that we had a positive incentive (laughing) to learn from the mistakes of others, and even our own. This fits with the benign violation theory of humor~\cite{Raskin2009} that says we find things funny when our expectations are violated in a way that is surprising, but not too threatening, disturbing, or wrong. This is why many jokes are insults that violate expectations about people's physical and mental well-being (insults), breaking social or cultural norms (rude behavior), or linguistic norms (word play). These jokes also help us sense what is unexpected, and learn from it.

Humor also has structure. Jokes typically have a setup that establishes expectations and a punchline that violates them. The semantic script theory of humor~\cite{Raskin2009} also describes how the setup leads listeners to infer one interpretation of the information, but the punchline leads them to infer a different interpretation that is also consistent with the facts (and the second interpretation of ``script'' often that violates a norm). Take this classic insult joke, \textit{"There are three types of people in the world. People who can count and people who can't."} The setup leads you to infer the person will list three types of people, but the punchline only lists two, which is unexpected. The second interpretation is that the person is bad at counting (insulting their own mental acuity). Constructing this joke required many cognitive skills beyond just knowing the setup-punchline structure - it required writing from someone's (mistaken) point of view, constructing a logical error someone could make, and finding a bland set-up line to misdirect listeners' attention to heighten the surprise. Beyond structure, it is unclear what is required of human or machine cognition to generate humor.

Although there is no definitive formula for humor, many humorists have described a set of techniques and skills for writing jokes. Key themes includes:

\textbf{Jokes should be relatable} ~\cite{Dean2000,Holloway2010, Vorhaus1994, Kaplan2013,Carter2001}, listeners have to related and empathize with them in some way.  
Jokes typically engage human emotions - fear, hope, curiosity, cringing, and other heightened states that raise the stakes that get us to listen and relate to the material  ~\cite{Carter2001}.  
Often jokes are for an in-group ~\cite{hurley2011inside} - using the shared knowledge and experiences of a group to create exclusive material which only that group could relate to.


\textbf{Jokes have details and observations}. Observing the difficulties and absurdities of life is a good way to find relatable material~\cite{Kaplan2013}.  But obvious absurdities tend not to be surprising, so jokes often come from observing details that others likely missed. ~\cite{Carter2001, Vorhaus1994}.

\textbf{Jokes contain narratives that include a point of view}.~\cite{Carter2001} 
Like stories and narratives, jokes often use multiple points of view to see a situation in a new and surprising way. 
Understanding peoples' thoughts, actions, and behaviors helps fully ``act out''~\cite{Carter2001} the story. 
For example, \textit{"Question: How do you get an Amish person to change a lightbulb? Answer: What's a lightbulb?"} this takes the Amish person's point of view in the answer, which is unexpected, the information is more immediate when it is told from the point of view of the person directly affected.
Sometimes that story can be based on a metaphor that helps the listener see something in a new way: \textit{"Playing an unamplified guitar is like strumming on a picnic table" (Dave Barry)} ~\cite{sep_humor}. Finding a good narrative, or point of view for a situation can help show the humor in it.

\textbf{Jokes are creative and can be intentionally constructed}
Almost all humorists who write books about humor
echo the design literature that divergent and convergent thinking are helpful processes in writing humor. 
Exploration is necessary to expand topics~\cite{Holloway2010} and many alternatives for punchlines should be explored to find the one that resonates the most~\cite{humor_Dean2000}.

Overall, jokes require complex human and social understanding. This is probably why humor difficult for machines to produce.


\subsection{Computational Creativity and Humor}

Computational humor is an outstanding challenge in AI. Even
classifying whether or not something is funny is a computational challenge. This has been tried with some limited success by training deep learning classifiers on New Yorker Captions ~\cite{shahafjokes}, and by "unfunning" jokes to create more training data~\cite{unfun}.
Wordplay is a common target for computational humor generation, and it relies on linguistic violations rather than social or cultural ones. Thus many successful systems have produced computational word play ~\cite{jape, twss, witscript1, he2019pungenerationsurprise, Taylor04computationallyrecognizing}, but these techniques don't extend to broader types of humor. 
Generating humor without wordplay is harder. 

LLMs have opened this as a possibility but humor continues to be a challenge for LLMs like ChatGPT~\cite{gptnotfunny, deepmind_humor}, even with fine tuning~\cite{anonymous2024funlms}. 
Recently, ChatGPT has been shown to be funnier than jokes written by Turkers (as rated by other Turkers), but not nearly as funny as professionally published humor ~\cite{gptvsturk}.

A common technique for non-wordplay humor is finding and relating unexpected associations ~\cite{witscript2}, using multiple humor strategies for associations~\cite{witscript3}.
Adding multi-step reasoning can also be applied~\cite{tikhonov2024humormechanicsadvancinghumor}, but is not sufficient. 
These attempts are largely in the right direction. Using associations, reasoning, and creative processes like divergent and convergent thinking made sense. But there are many more dimensions to humor. Considering more human-centric ideas like point of view, observation of details, social understanding, and understanding the audience is also essential for humor generation. 


\subsection{AI Generation Techniques}

Prompt engineering is the most general AI generation technique~\cite{gpt2}. Crafting prompts that clearly define the problem and the expected output is the first technique to try to generate better outputs. However, when prompt engineering is insufficient there are many other generation paradigms that have been shown to achieve better results.
Fine-tuning allows the model to learn the specific patterns of a target output type. This can help learn writing style, topics, and other implicit aspects of language.
Results from a model can be chained together to solve a bigger problem~\cite{cai_ai_chains}, allowing the model to focus on solving one problem at a time. 
Chain of thought~\cite{cot} is a technique that helps models get more accurate answers to math questions by talking through the steps. It is somewhat equivalent to the human technique of "showing one's work". 
Similarly, thought experiments can better solve moral questions by thinking through the consequences of possible actions before making judgments ~\cite{ma2023letsthoughtexperimentusing}.
Reflection ~\cite{reflection} is a technique for AIs to evaluate their own outputs, and try to improve based on their own evaluation. 
Using AI to evaluate its own outputs, does not always produce good results, but it is commonly used~\cite{ai_self_eval}. 
Agents~\cite{joon_agents} can be useful to have AIs talk to one another to reason through a problem. Agents of different skills (or roles) can collaborate to do hard tasks that require multiple perspectives or roles like software engineering, and solving physics problems. A mixture of experts or perspectives can also be used to improve results ~\cite{cai2024surveymixtureexperts}.


In our system, we use fine-tuning to give the LLM a better sense of the target artifact. Implicit in this is the tone, the style, and the vocabulary expected in the humor. 
We use chains to separate stages of the humor generation process. We have an observation stage that makes implication information in images explicit, similar to the spirit of chain-of-thought and thought experiments. We also have an LLM as an evaluation of outputs - using a fine-tuned LLM trained to evaluate humor specifically for the target audience, somewhat like a mixture of experts model.
There are several other AI paradigms that we did not explore here, but could be relevant to humor including: reflection planning, question asking, retrieval augments generation (RAG), and human reinforcement learning.

\section{System}
\begin{figure*}[h]
    \centering
    \includegraphics[width=.85\textwidth]{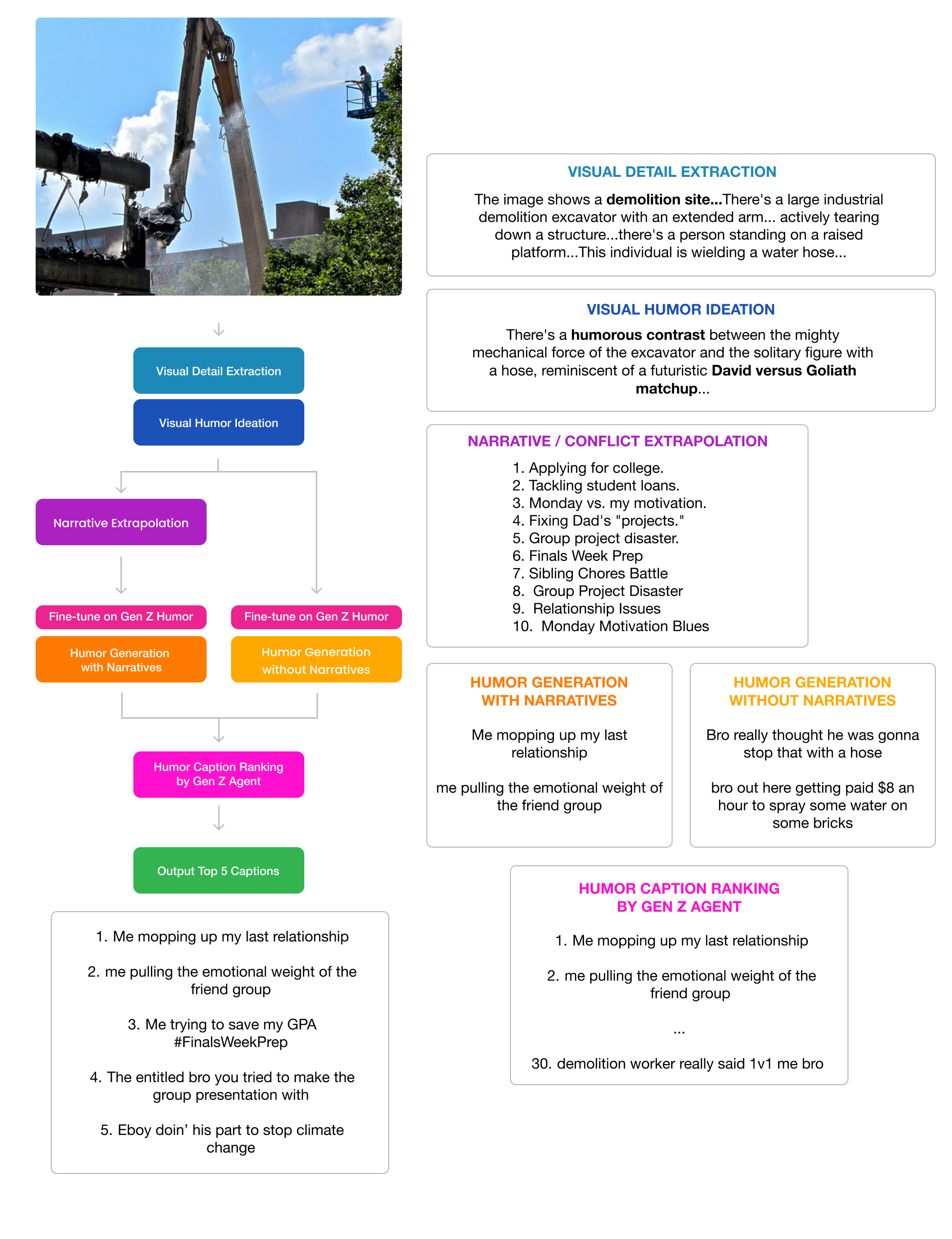}
    \caption{HumorSkills System Diagram. Given an image, the system first extracts visual details with a visual language model, then performs visual humor ideation to analyze the image and propose humorous angles. It then generates ten potential conflicts that could be used to extrapolate the image into a relatable experience. The system then generates humor with and without the narratives, for diversity. A separate instance of the LLM trained to rank gen-Z humor ranks all the captions and returns the top five.}
    \Description{HumorSkills System Diagram}
    \label{fig:system}
\end{figure*}

HumorSkills is a system that takes an input image and outputs 5 image captions. 
The architecture has three key steps that mimic human skills needed for humor. \textit{Visual Detail Extraction}, is a step that describes the image in depth in order to make non-obvious observations about it. \textit{Narrative and Conflict Extrapolation} is a step that finds narratives not in the image that could be related to it, to expand the topic of jokes to things that are not just in the image but also analogous to it.  \textit{Fine-tuning} the joke generator with examples of good Gen-Z humor helps the jokes be more relatable to the target audience by using references, slang, topics, and insecurities that resonate with this group.


The system generates two types of captions: image-focused captions which common directly on the content in the image, and narrative-driven captions. Variety is important to humor. Humor relies on surprise, and jokes that are too similar start to become more predictable. Additionally, with an infinite set of input images with different subjects and situations, there are more strategies needed to find a humorous angle that fits the content. 



\subsection{AI Humor Generation Walk Through}
Figure \ref{fig:system} contains a visual diagram and example of intermediate outputs when generating captions for an image. We describe each phase and implementation in detail.  

\subsubsection{Visual Detail Extraction}

The first phase of the system’s workflow involves the Visual Detail Extraction component, which utilizes GPT-4o’s vision capabilities to analyze the input image. This system incorporates a prompt that asks for a detailed paragraph that explains the who/what/where of the image, distinguishing between identifying the subject of the image, the main action of the image if it exists, and the background elements of the image. This component is responsible for extracting key visual elements such as objects, human expressions, background settings, and any notable aspects that could serve as the foundation for humor.

For instance, in the demolition site example from the system diagram, the system identifies a large industrial demolition excavator and a person with a hose spraying the demolition site. 

\subsubsection{Visual Humor Ideation}

On top of the visual detail extraction, the system ideates on possible humorous elements from the visual of the image. This incorporates an additional prompt using GPT-4o that intakes the image and asks it to identify and ideate on potential humorous visual elements in the image, whether they are directly humorous elements, such as funny facial expressions, or more analogous elements. For example, for the system diagram image, the system noted the visual contrast of the excavator and person, reminiscent of a David versus Goliath scenario, which provides a foundational metaphor for generating humorous captions. 

\begin{figure*}[b]
    \centering
    \includegraphics[width=.95\textwidth]{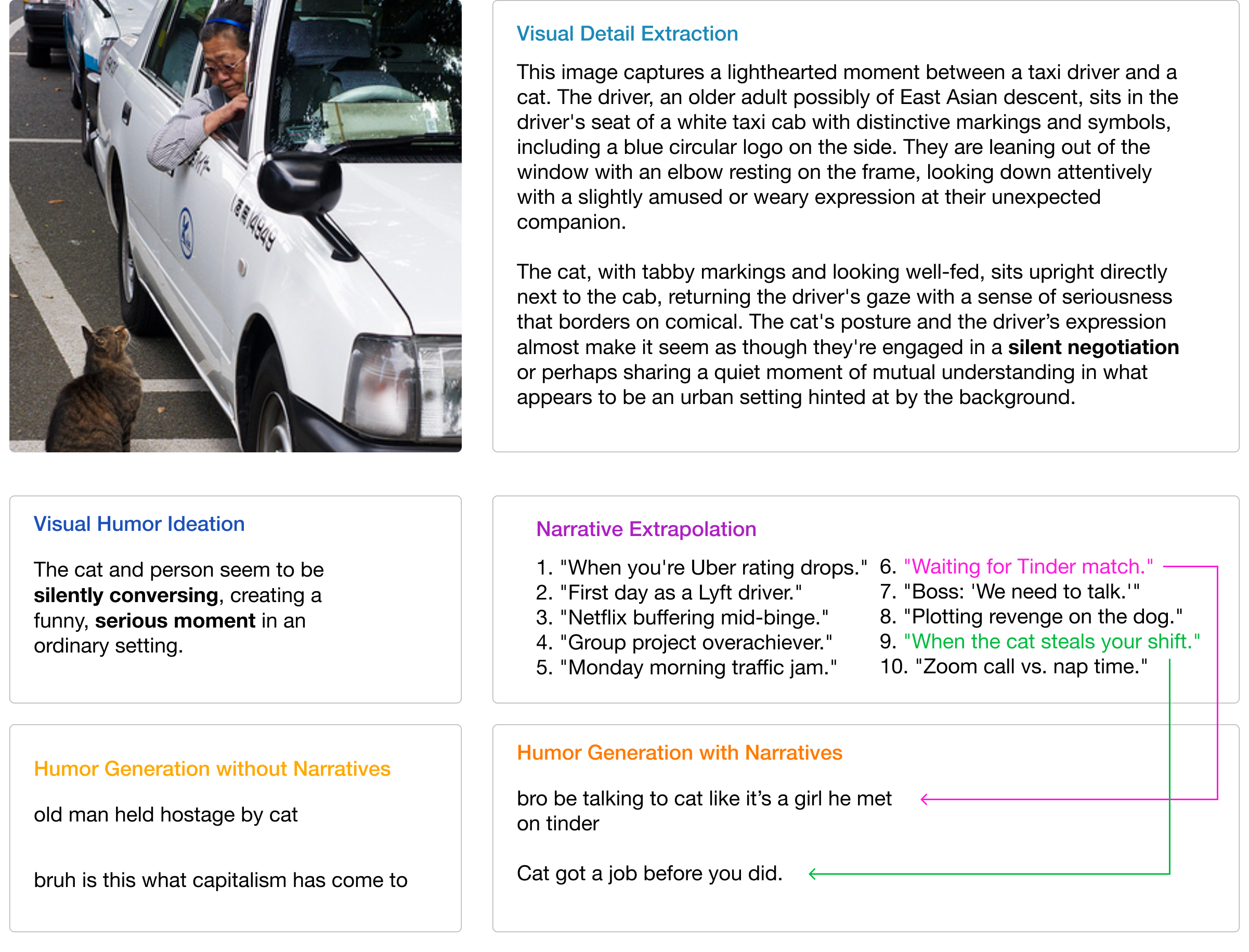}
    \caption{A diagram for how narrative extrapolation works}
    \label{fig:systemLines}
\end{figure*}

\subsubsection{Narrative and Conflict Extrapolation}

In this next step, the system generates a narrative and conflict framework by drawing upon common and relatable Gen Z experiences such as work, school, family, social interactions, relationships, and more. 
The system chains together the results of the previous steps, into a new prompt sent to GPT-4o. 
The prompt contains the visual details, the visual humor ideation, and a list of common Gen Z experiences,  and the instruction to "generate narratives that reflect the essence of the image that is set within the framework of the Gen Z experience."
This narrative generation adds depth to the humorous captions by applying relatable themes and conflicts to the visual elements identified earlier.

For instance, our system diagram generates narratives such as “Tackling student loans”, "Group Project Disaster", and “Relationship Issues” based on the image, both of which are common experiences among those who identify as Gen Z. These particular narratives are likely inspired by the imagery of a disaster site, referring to how the effort of paying off student loans, attempting to complete group projects during school, or addressing relationship -- all of which can feel like disaster clean up. These relatable conflicts can transform the visual of a demolition scene -- a setting that is not particularly relatable -- into a relatable scenario that has the potential for humor, thereby expanding the humorous possibilities by connecting the visual input with broader life experiences.

\subsubsection{Humorous Caption Generation}

Following the narrative and conflict extrapolation, the system generates humorous captions in the generation stage using a fine-tuned version of GPT-3.5 trained on humorous Instagram comments. This involves producing captions through two distinct strategies: one focused on the visual humor of the image, and the other by bringing in the previously generated external narratives. Caption generation is segmented into two separate prompts utilizing the fine-tuned GPT-3.5 model. For captions without generated external narratives, the prompt asks to generate 15 humorous captions in the style of Gen Z that bases the generation off the visual extraction and visual humor ideation of the input image. For captions with the external narratives, the prompt also asks to generate 15 humorous captions in the style of Gen Z that bases the generation off the visual extraction and visual humor ideation of the input image, but also asks the system to incorporate the list of generated narrative/conflict extrapolations to base the humorous captions off of.

Image-focused captions rely solely on the visual details in the image, such as “bro out here getting paid \textdollar8 an hour to spray some water on some bricks,” which references the direct visual elements in the scene in order to generate a caption. This particular caption directly references the humor of the image, poking fun at the minimal impact of the person spraying water on bricks while an excavator clearly has more impact on the demolition site. Narrative-driven captions, on the other hand, introduce external references to add humor. For instance, a caption like “The entitled bro you tried to make the group presentation with” introduces an outside, exaggerated, interpretation of the scene from earlier, "Group Project Disaster." This caption takes the group project narrative and pairs it with the visual of the image, analogizing the person spraying the hose with minimal impact on the demolition site to an entitled person who has not done much to complete the group project. 

This variety between visual humor and narrative-driven humor is crucial because jokes that are too similar become predictable, losing their element of surprise. Additionally, humor strategies need to adapt to the varying content in input images. Some images lend themselves to humor based on their inherent visual details, while others require bringing in outside references to create a joke. For instance, an image of static objects might not be inherently funny, such as the demolition image shown in the system diagram, but a caption introducing an unrelated, exaggerated narrative, such as “Eboy doin' his part to stop climate change” can inject humor and absurdity by making an unexpected connection.

\subsubsection{Caption Ranking using Gen Z Agent}

The final component of the system architecture is the Caption Ranking and Filtering Agent, a GPT-4o-based agent fine-tuned to evaluate humor from a Gen Z perspective. This agent receives the list of 30 total captions from the narrative and visual humor-based caption generations and ranks the captions generated in the previous stage based on humor, relatability, and alignment with the image and narrative.

As illustrated in our system diagram, this agent ranks captions such as “Me mopping up my last relationship” and “me pulling the emotional weight of the friend group” based on their relevance to Gen Z humor. Captions that fail to meet the humor threshold are filtered out, such as "Demolition worker really said 1v1 me bro," because although the phrase like "1v1 me bro" invokes Gen Z phrases, the content of the caption seems less relevant and relatable than a caption talking about school or relationships, ensuring that only the most effective and relatable captions are presented to the user.

\subsubsection{Fine-tuning}

To fine-tune a GPT-3.5 model, a dataset of 80 humorous comments were extracted from popular Instagram images. From three popular Instagram meme pages with over 400,000 followers, the top five comments of each image post were collected. All fit the style of Gen-Z humor. 
Examples of the visual description of the images in addition to an explanation of potential humorous elements of the image were written in the fine-tuning prompts, then followed by the actual comment itself. This reflected the visual extraction and humor ideation being incorporated into the prompt of our current system.


\section{Humor Rating Study}
\subsection{Methods}
We ask people to rate humorous captions humor for images taken from 8 popular Instagram (IG) humor captioning accounts. For each image, users were shown 15 captions
1) The top 5 most upvotes captions from IG
2) 5 captions from GPT-4o
3) 5 captions from HumorSkills. 

Data was collected from an online survey. Users would see an image and then rate 15 captions for it on a scale of 1 to 5 where 1 was "not funny", 3 was "somewhat funny" and 5 was "very funny". For each user, the images were presented in a random order and the 15 captions for each image were also shown in a random order to avoid any possible ordering effects. Users were not told the condition of each caption. 

The survey was distributed through email announcements to clubs and classes at a local university. Users were not told that AI was involved in the humor writing. The announcements advertised as being for people who identify as "Gen-Z" and who "like Instagram caption humor". These qualifications were added to ensure the raters were part of the target audience, and most university students fit this demographic.  The survey took about 30 minutes and users were paid \$10.

Our Hypotheses were that:

H1) Our system would be rated as funnier than GPT-4o, but that 

H2) Our system would be rated as funny as the top-rated Instagram captions.

\subsection{Results}
In total, 32 people responded to the survey, 5 male, 9 female, and 18 declined to say their gender. 14 of 32 respondents were aged 18-25. 1 were aged "25+", and 17 declined to say their age group.

To test the relative humor score of the three conditions, we ran a Generalized Linear Mixed Model (GLMM) to estimate the funniness rating of each condition. The response types were ordinal numbers (ratings), the fixed effects were the caption type (IG, GPT-4o, System), and the random effects were individual rater IDs and image IDs.

\subsubsection{H1: HumorSkills vs. GPT-4o}
On the whole, captions from our system were rated as 2.27 (out of 5) for funniness. Captions written by GPT-4o were rated less funny (by 0.213 points), which is statistically significant at a p<0.0001 level. This indicates our system writes more humorous captions than a state-of-the-art VLM with prompt engineering. 
Results are shown in figure \ref{glmm_1}. 
Thus \textbf{H1 is supported. Our system is funnier than GPT-4o.}

\subsubsection{H2: HumorSkills vs. top IG captions}
On the whole, the top IG captions were rated only slightly better (0.08 points on a 5-point scale).
But the difference was not significantly different at the 5\% level. However, it was very close (p=0.053)
This indicates that the system is competitive with the top IG captions. This effect could disappear with more data, but this dataset had over 6000 observations, and the 0.08 difference in average score is only 2\% of the 1-5 scale.
Thus \textbf{H2 is supported. Based on our sample, our system was rated as funny as the top 5 Instagram comments.} Or more precisely, our system was not statistically less funny than the top IG captions. 

\begin{table}
\caption{Mixed Linear Model Regression Results}
\label{glmm_1}
\begin{center}
\begin{tabular}{llll}
\hline
Model:            & MixedLM & Dependent Variable: & rating       \\
No. Observations: & 6015    & Method:             & REML         \\
No. Groups:       & 402     & Scale:              & 1.6212       \\
Min. group size:  & 2       & Log-Likelihood:     & -10271.0088  \\
Max. group size:  & 15      & Converged:          & Yes          \\
Mean group size:  & 15.0    &                     &              \\
\hline
\end{tabular}
\end{center}

\begin{center}
\begin{tabular}{lrrrrrr}
\hline
                        &  Coef. & Std.Err. &      z & P$> |$z$|$ & [0.025 & 0.975]  \\
\hline
Intercept (HumorSkill rating)             &  2.273 &    0.040 & 56.601 &       0.000 &  2.194 &  2.351  \\
GPT-4o rating & -0.213 &    0.040 & -5.285 &       0.000 & -0.291 & -0.134  \\
Top5 IG rating &  0.078 &    0.040 &  1.934 &       0.053 & -0.001 &  0.157  \\
Group Var               &  0.323 &    0.025 &        &             &        &         \\
\hline
\end{tabular}
\end{center}
\end{table}

\subsection{HumorSkills performance on non-target images}
The HumorSkills approach was designed and fine-tuned on a particular type of humor - Instagram captioning. It is possible that this approach is not generalizable, and too well tailored to one domain. The Instagram Caption Challenge is selected by the posters for particular qualities, which aren't stated, but a consistent theme is that the images are already interesting and unusual.  They almost always include people interacting. However, other types of images could be harder to caption humorously. We test two other sources of images to see how well HumorSkills could caption those images compared to GPT-4o.

1. \textbf{Camera roll images} -- the everyday photos people take -- typically contain people, but the images are not necessarily interesting in their subject or composition. 

2. \textbf{Museum Art}. Art is very different than photos. Although some of it depicts people doing things. Most of those things are historical, and not particularly relatable. Some paintings are abstract, like Jackson Pollock's, do not concretely depict any discernible subject. Additionally, many of the art pieces are not paintings but objects like vases or chairs. Any image without people are challenging to make relatable and create a narrative for.

We randomly selected 30 images from the Flickr Image Dataset\cite{young2014image} and 30 museum art images to test on our target audience. We take 15 images from the Museum of Modern Art (The MoMA) and 15 images of art from from the Metropolitan Art Museum (The Met). For every image, we generate 5 HumorSkills captions and 5 GPT-4o captions. We create two separate surveys for rating the captions - one for art and only for camera roll images. As before, we randomize the order of the images and the captions to mitigate ordering effects. We recruited raters the same as for the previous study.

We hypothesize that HumorSkills' captions will be rated more funny than GPT-4o captions. We did not compare to top-ranked human-written captions because these images were not taken from Instagram.

\subsection{Results: Camera Roll Humor Ratings}
As before, we ran a GLMM regression with image ID and user ID as random effects. There were a total of 4765 ratings. There were a total of 21 respondents, there were a total of 10 female, 4 male, and 7 declined to say their gender. 12 respondents were aged 18-25, 2 were aged "25+", and 7 declined to say their age group.

The results show that HumorSkills is 0.291 points higher than GPT-4o (2.19 vs. 1.9 out of 5). This difference is statistically significantly higher at the 0.05 level (p=0.022). Results are shown in Table \ref{glmm_flickr}.

\begin{table}
\caption{FlickrImage GLMM Regression Results}
\label{glmm_flickr}
\begin{center}
\begin{tabular}{llll}
\hline
Model:            & MixedLM & Dependent Variable: & Rating       \\
No. Observations: & 4765    & Method:             & REML         \\
No. Groups:       & 30      & Scale:              & 0.5792       \\
Min. group size:  & 149     & Log-Likelihood:     & -6209.2910   \\
Max. group size:  & 170     & Converged:          & Yes          \\
Mean group size:  & 158.8   &                     &              \\
\hline
\end{tabular}

\vspace{0.5cm}

\begin{tabular}{lcccccc}
\hline
\textbf{Variable} & \textbf{Coef.} & \textbf{Std.Err.} & \textbf{z} & \textbf{P>|z|} & \textbf{[0.025} & \textbf{0.975]} \\
\hline
Intercept (GPT-4o rating)      & 1.901  & 0.054  & 35.357 & 0.000 & 1.796 & 2.006 \\
HumorSkill rating   & 0.291  & 0.022  & 13.207 & 0.000 & 0.248 & 0.334 \\
Group Var      & 0.000  &        &        &       &       &       \\
User\_ID Var   & 1.260  & 0.116  &        &       &       &       \\
\hline
\end{tabular}
\end{center}
\end{table}

\subsection{Results: Museum Art Humor Ratings}
There were a total of 4044 ratings. There were a total of 17 respondents, 7 female, 3 male, and 7 declined to say their gender. 8 respondents were aged 18-25, 2 were "25+", and 7 declined to say their age group.

The results show that HumorSkills is 0.18 points higher than GPT-4o (2.18 vs. 2.00 out of 5). This difference is statistically significantly higher at the 0.05 level (p=0.023). Results are shown in Table \ref{glmm_art}. This shows that HumorSkills can perform out-of-domain images as these images do not have obvious humorous qualities to them. Nonetheless, HumorSkills still apply.

\begin{table}
\caption{Art Image GLMM Regression Results}
\label{glmm_art}
\begin{tabular}{llll}
\hline
Model:            & MixedLM & Dependent Variable: & Rating       \\
No. Observations: & 4044    & Method:             & REML         \\
No. Groups:       & 30      & Scale:              & 0.5241       \\
Min. group size:  & 129     & Log-Likelihood:     & -5106.3412   \\
Max. group size:  & 150     & Converged:          & Yes          \\
Mean group size:  & 134.8   &                     &              \\
\hline
\end{tabular}

\vspace{0.5cm}

\begin{tabular}{lcccccc}
\hline
\textbf{Variable} & \textbf{Coef.} & \textbf{Std.Err.} & \textbf{z} & \textbf{P>|z|} & \textbf{[0.025} & \textbf{0.975]} \\
\hline
Intercept (GPT-4o rating)      & 2.000  & 0.061  & 32.774 & 0.000 & 1.881 & 2.120 \\
HumorSkill rating   & 0.175  & 0.023  & 7.677  & 0.000 & 0.130 & 0.219 \\
Group Var      & 0.001  &        &        &       &       &       \\
User\_ID Var   & 1.387  & 0.147  &        &       &       &       \\
\hline
\end{tabular}
\end{table}

\begin{figure}
    \centering
    \includegraphics[width=.8\textwidth]{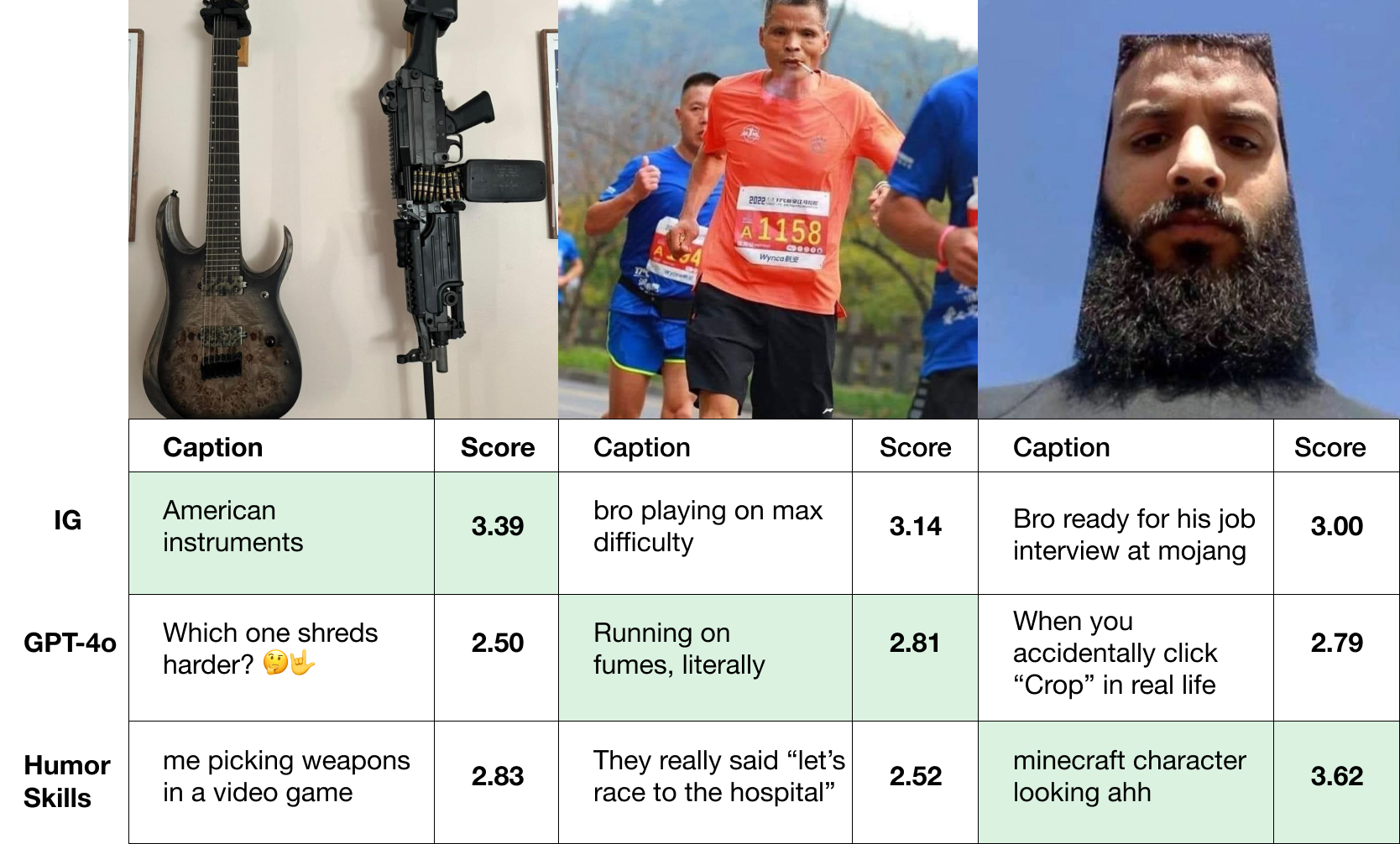}
    \caption{Top Rated Image Captions (marked in green) for Instagram, GPT-4o, and HumorSkills and corresponding top scorers for each image. From left to right, the images contain 1) a guitar next to a machine gun hanging on a wall, 2) a man running a race while smoking a cigarette and, 3) a man with a long beard and his head cropped to a trapezoid shape.}
    \label{top_score}
\end{figure}

\section{Results: Qualitative Analysis}

\subsection{Top Rated Images for Instagram, 4o, and HumorSkills for Target Images}

All three generation techniques were able to produce captions that some people found funny. Figure \ref{top_score} shows the top-rated captions for each of the three techniques. 
The top-rated IG comment is “American instruments,” which pokes fun at US gun culture. This caption is good because it's short but also an excellent social critique. The GPT-4o caption for this image is also apt: “Which one shreds harder?” It understands the image and associates the two objects through word play ("shred"). The HumorSkills caption for this image draws an analogy to video games. It's not as sharp a critique as the IG captions, but it's very relatable and critiques the silly weapons available in many video games. Although this association might not be clear to everyone, it will be to Gen-Z.


The top-rated GPT-4o caption says "running on fumes, literally", for an image with a runner smoking a cigarette. This is apt and connects the odd parts of the image (running and smoking). It again uses wordplay. 
The IG caption was the highest rating for this image, with an unexpected video game reference of ``bro playing on max difficulty.''
The HumorSkills caption did least well for this image, it uses dark humor to critique the runner for smoking and point out the absurdity: “They really said ‘let’s race to the hospital.’” 


The top-rated HumorSkills caption was the highest-rated caption in the entire study. It reads ``minecraft character looking ahh.'' (The word "ahh'" is Gen-Z slang roughly meaning "ass".) Although this might not be broadly funny, it is funny to Gen-Z because of this insider slang, because it's short, and because it is a critique of his looks related to Minecraft. In the online Minecraft videogame, characters are made out of blocks, like virtual legos, and their heads are often square, which is reminiscent of how the face in this photo to cropped.
The IG caption (``bro ready for his job interview at mojang'') also relates the image to Minecraft (Mojang the developer of Minecraft). Gen-Z grew up on Minecraft, and jokes about it resonate. The GPT-4o also makes fun of his hairstyle but relates it to cropping an image, rather than Minecraft, which is not as resonant. 


\begin{figure*}[h]
    \centering
    \includegraphics[width=.95\textwidth]{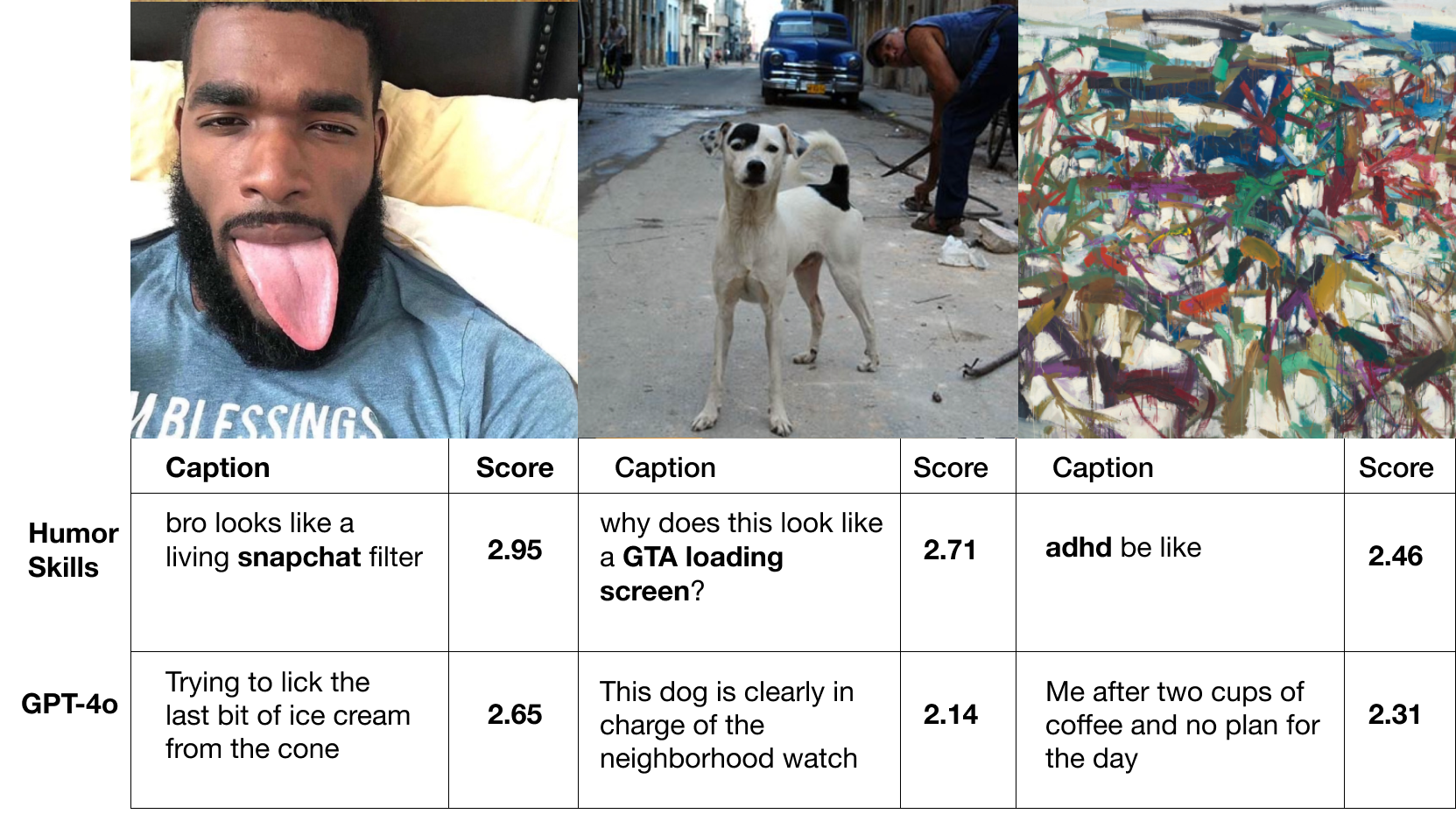}
    \caption{Images across all 3 datasets with Gen Z slang. Instagram (left), Flickr (center), Museum Art (right)}
    \label{fig:visual_images}
\end{figure*}

\subsection{Gen Z Slang and Fine-Tuning}
Figure \ref{fig:visual_images} shows three images from across all three datasets. Although the HumorSkills captions are rated slightly better than the GPT-4o jokes, they are fairly good.

Continuing with examples from the Instagram dataset shows that HumorSkills can easily replicate Gen Z slang through its fine-tuning. The first image, featuring a man with a long tongue, has the generated HumorSkills caption “bro looks like a living snapchat filter.” This caption plays on the exaggerated feature of his tongue — and ties it directly to Snapchat. Snapchat filters are an embedded part of Gen Z’s social media use, making the caption incorporate phrases familiar to Gen Z, such as Snapchat. This is ignored in the GPT-4o caption, "Trying to lick the last bit of ice cream from the cone." While eating ice cream is commonplace for many, Snapchat is a phrase more unique to Gen Z.

Even in a non-target dataset, such as the Flickr dataset, HumorSkills’ ability to produce Gen Z slang can be seen in the second image, displaying a dog standing defiantly in the middle of an urban street. The caption, “why does this look like a GTA loading screen?” cleverly references Grand Theft Auto (GTA), a highly popular action-adventure game among Gen Z. In GTA, loading screens often feature characters and animals in striking, dramatic poses against the backdrop of the game’s open-world urban environment. These screens typically display gritty street scenes filled with elements of crime, defiance, and tension, mirroring the tone of the game itself.

The image of the street dog captures this same defiant, street-wise aesthetic that players recognize from the game, evoking the vibe of the cityscapes and characters GTA is known for. HumorSkills recognizes this visual parallel and ties it to a cultural touchstone that is deeply ingrained in Gen Z’s gaming experiences. The phrase “GTA loading screen” instantly brings to mind the iconic visuals and rebellious energy of the game, enhancing the humor through a shared cultural reference and common phrase popular among Gen Z. On the other hand, the GPT-4o caption, "This dog is clearly in charge of the neighborhood," while relevant to the dog and urban environment, does not contain a cultural or popular phrase among Gen Z like GTA.

In the museum art dataset, for more abstract images, such as the third image with bold, chaotic strokes and splashes of color scattered across the canvas, the caption, “adhd be like,” uses the chaotic and unfocused nature of the painting as a metaphor for adhd. This humor also resonates with Gen Z, who are more comfortable discussing mental health and will throw around the term "adhd" more. The abstract painting lends itself to this caption, as it reflects the disordered, busy, and scattered hardship associated with the condition that Gen Z mentions a lot. The GPT-4o caption, "Me after two cups of coffee and no plan for the day," also addresses a similar condition of feeling chaotic after drinking coffee, but does not invoke a relatable condition relating to mental health like "adhd" that is popular among Gen Z.

\begin{figure*}[h]
    \centering
    \includegraphics[width=.95\textwidth]{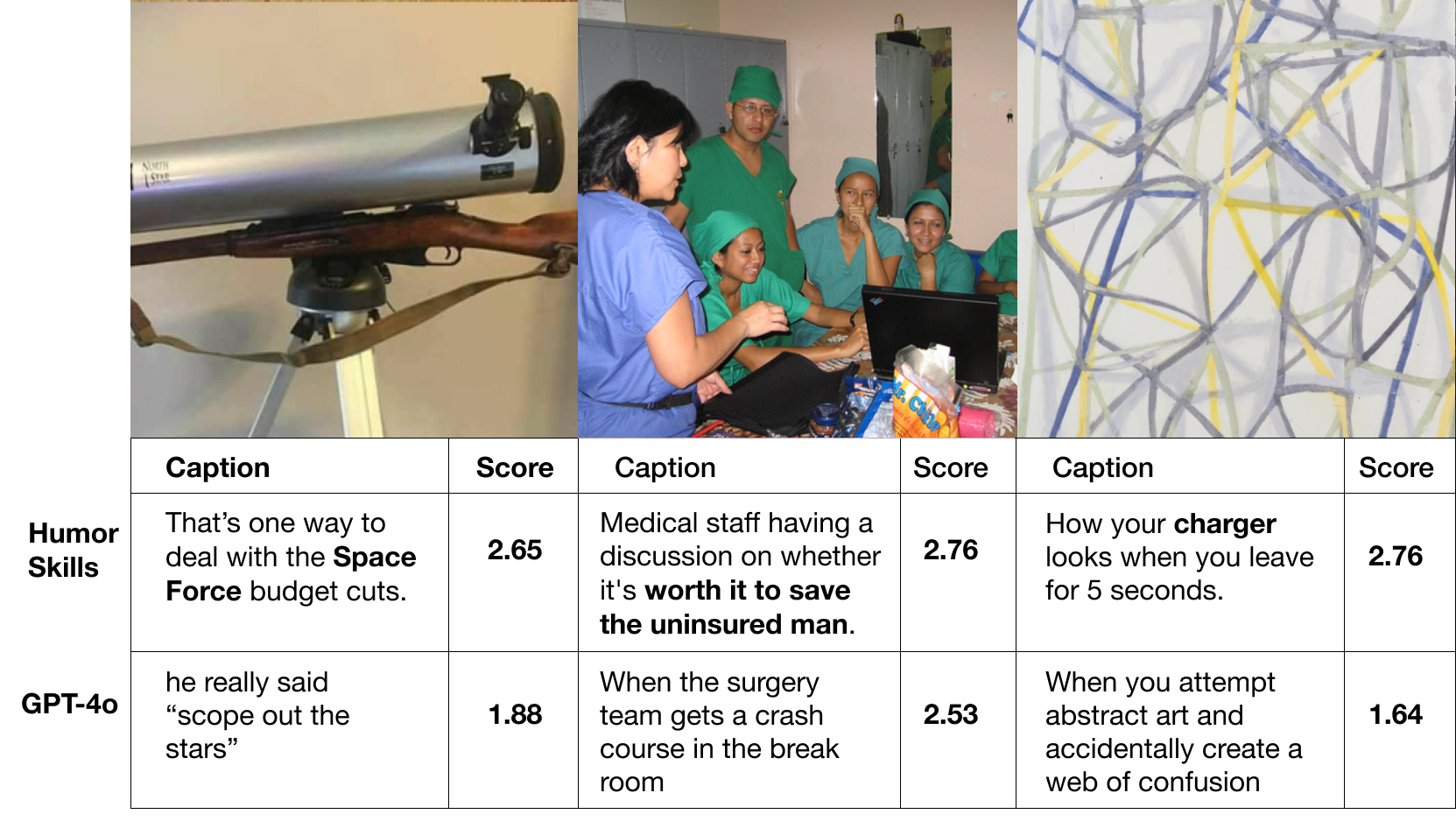}
    \caption{Images across all 3 datasets with narrative generation. Instagram (left), Flickr (center), Museum Art (right)}
    \label{fig:narratives}
\end{figure*}

\subsection{Using Narrative Extrapolation for Relatability}
While humor generation with Gen Z slang seems natural on images HumorSkills was built for and trained on, HumorSkills can generate narratives for images where it is not obvious there is Gen Z slang available to produce into a humorous caption. 

The first image from Instagram shows a comically large telescope on top of a rifle as if it were a sniper scope. With no apparent narrative for the image, HumorSkills generates one by assuming the event in which the US Space Force has faced budget cuts with the caption, "That's one way to deal with the Space Force budget cuts," implying that the telescope, used in space, and the rifle, used by the military, has been jury-rigged together for use by the US Space Force due to budget constraints. Without the narrative generation, GPT-4o produces a relevant caption about the telescope, "he really said 'scope out the stars'," but it does not contain a narrative or core conflict to apply to the image. 

The second image shows a group of medical professionals gathered around a laptop, seemingly discussing something serious. The caption, “Medical staff having a discussion on whether it’s worth it to save the uninsured man,” comments on the state of healthcare, particularly in countries like the United States, where lack of insurance can impact the quality of care. This caption utilizes narratives to insert a relatable critique of the healthcare system, framing the scene as if the doctors are debating the value of a human life based on their insurance status. The humor here comes from how HumorSkills can create an extreme suggestion using a narrative play on a real-world issue many can relate to the cost of healthcare. The GPT-4o caption, "When the surgery team gets a crash course in the break room," while addressing the humorous nature of expert surgeons receiving lessons while on break - suggesting the incompetency of the surgeons, does not establish a coherent narrative with a clear conflict like saving an uninsured man like in the HumorSkills example. Rather, the issue is that the images imply the surgeons are incompetent (who are all in the image) rather than another character (the man) with a conflict (he is uninsured). 

The third image shows an abstract painting with a web of intersecting lines. The caption, "How your charger looks when you leave for 5 seconds," utilizes a narrative of getting up after using an electronic device such as a mobile phone or computer with a charger attached, only to come back to find it twisted and tangled. While there are no physical wires in the painting, HumorSkills generates a hypothetical narrative as if the lines are wires attached to a charger. The GPT-4o caption, "When you attempt abstract art and accidentally create a web of confusion," while also a narrative, pokes fun at the attempt of the actual painter trying to create abstract art only for it to be confusing. The HumorSkills caption is unique in that the narrative is separate from the actual image by generating a scenario of the painting representing someone's charger wires for their devices. The GPT-4o narrative focuses on a narrative for the actual image itself and the identity of its painter.

\begin{figure*}[h]
    \centering
    \includegraphics[width=.95\textwidth]{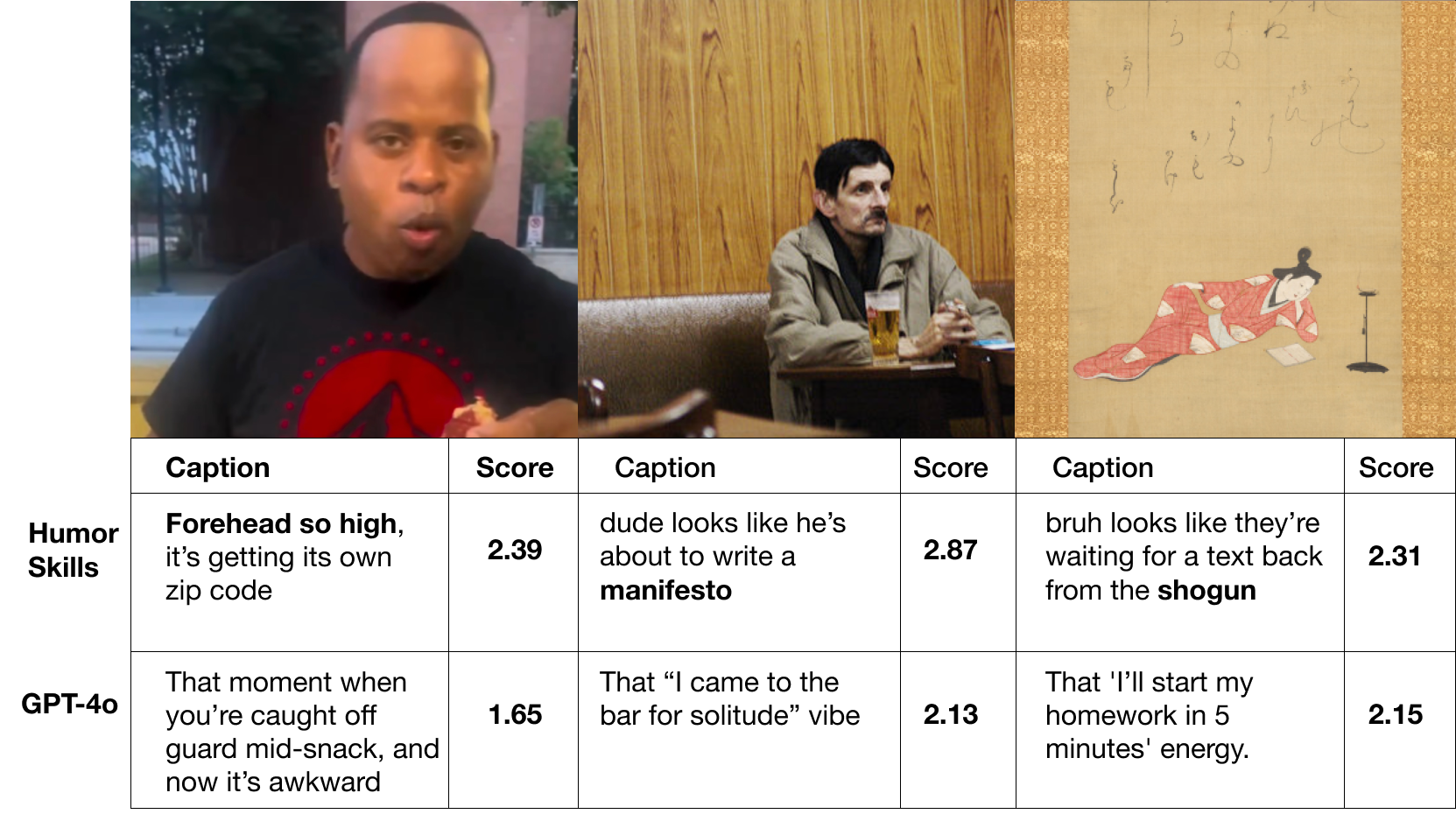}
    \caption{Images across all 3 datasets with notable visual extractions. Instagram (left), Flickr (center), Museum Art (right)}
    \label{fig:visual_system}
\end{figure*}

\subsection{Visual Detail Extraction and Visual Humor Ideation}

In the first image, we see a man eating food with a large forehead. The caption, "Foreheard so high, it's getting its own zip code," refers to the abnormally large forehead of the man, that the height of the forehead is so far that it warrants its own area code. While GPT-4o most definitely can pick up on this visual abnormality, it often drifts towards non-significant visual details and fixates on them, as seen in the caption, "That moment when you're caught off guard mid-snack, and now it's awkward," fixing on the man-eating food while being photographed while chewing. GPT-4o definitely will pick up on the man's forehead, but because it is not looking for visual oddities that may be considered funny (visual humor ideation) or prioritizing visual elements to the scene, it often reverts to a less prominent visual element. 

In the second image, we see a man with a mustache sitting alone with a beer. The caption, “dude looks like he’s about to write a manifesto,” plays on the stereotype of a brooding, isolated man, on the verge of drafting some kind of extreme ideological statement like many extremist historical figures. This is an example of visual extraction, where the system uses the man’s appearance of the mustache, brooding imagery, longing face, and demeanor to label an assumption about the man's political beliefs. The GPT-4o image caption, "That 'I came to the bar for solitude' vibe" directly addresses the solitude nature of the man in the photo without addressing the brooding emotion of the man, the mustache, and other elements like in the HumorSkills caption that helps assume a political ideology or belief system.

In the third image, it is a painting of a Japanese woman in a kimono lying flat reading a book. HumorSkills generated, "bruh looks like they're waiting for a text back from the shogun," highlighting that HumorSkills picks up the cultural visuals of the image from Japan by referencing the term shogun. While GPT-4o's caption is relatable with, "That 'I'll start my homework in 5 minutes' energy," it lacks the visual cue of a Japanese artwork and misses mentioning it in the image.

\section{Discussion}
\subsection{Giving Human-like Skills to AI} 
This study showed that for one form of humor - Gen-Z style Instagram image captioning humor - our AI-written humor was funnier than GPT's native humorsense, and as funny at the top 5 highest rated Instagram captions. We attribute this to a variety of features we added to the system. First, the visual detail extraction was able to find aspects of the image to poke fun at that were often sharper than GPT's joke target and more similar to the Instagram captions' joke target. Second, the narrative extrapolation step allowed the system to broaden its base of relatable joke targets - moving the focus away from making fun of the literal objects in the image, but using them as metaphors for relatable joke targets like relationship disasters, teamwork breakdowns, and the burden of student loans. This opened more creativity possibilities for joke targets. Lastly, we used an LLM-as-judge to rank the outputs accord to Gen-Z humor taste, thus giving the system some notion of the audience. These skills - detail observation, finding analogous and relatable social situations, and modeling the audience through fine tuning - are all considered somewhat ``human.'' Skills like reasoning and chaining are considered more typical of machines. But this shows that machines might be able to approach these more human skills with the right architecture and training.


There are many ways to improve the skills in HumorSkills. Building and testing a better Gen Z humor ranking would probably improve the filtering of bad captions. More fine-tuning could improve the breadth of Gen Z slang and references. More narratives and conflicts would expand its vocabulary of relatable situations. Finding ways to automatically collect narratives and conflicts to be applied would accelerate this process. Adding new skills would also be future research. Theories of humor abound. With recent advances in LLM's ability to do long chains of logical reasoning in DeepSeek and GPT-4o, it would be interesting to have AI try to analyze the humor and extract it's own theories or techniques for humor.

One of biggest shortcomings of the captions is that some of them are not logical enough to make sense, but are also not illogical enough to be absurd. These sound like mistakes. As future work, one could test whether an AI-based reflection step could think through the logic of a joke and decide whether it actually made sense or not. 


Further testing or ablation studies could help shed light on which skills are most helpful. However, humor ratings have high variance among raters, and the data required to get statistical significance is often quite high. There may not be an effect of each skill individually - they might only work together.

\subsection{Implications of Machine with Human-like Social Skills}
Human-like social skills - like humor - are often used for human bonding. If AI can write humor as well as the best people, the AI has the potential to both disingenuously create human bonding ~\cite{diresta2024spammersscammersleverageaigenerated,naaman_opinion} and to augment human's ability to bond~\cite{socialglue}. Either way, this has the potential to change the nature of human trust and communication.
In many ways, this is already happening in other domains. 
ChatGPT and Gmail SmartCompose~\cite{smartcompose} can already rewrite emails to sound more polite and we really are.
AI sales and scams can trick people into giving money to what they think are friend or loved ones in need~\cite{ai_scams}. 
AI has successfully been integrated into Gen Z dating apps that suggest messages to send to potential dates based on both a dating profile (for the opening line) or message history (for continued conversation)~\cite{majic2024rizz}. Many apps attempt this, but the quality of the suggested text sets them apart - the apps that generate more human-like texts have millions of active paying users~\cite{majic2024rizz}.
To some, this potential for disingenuousness is horrifying. Although disingenuous portrayals of oneself for dating purposes far precede the invention of generative AI, there is a possibility that AI will amplify this ability.

As AI for social, cultural, and personally relevant communication improves, we may need a way to discern genuine from disingenuous communication. There are high-tech ways of doing this, such as making a video of oneself (until AI can do that). There are also low-tech ways of doing this, like talking in person. It would be highly ironic if the advancement of AI drives people to abandon technology, because it could not be trusted to be genuine. 






\section{Limitations}
This study targeted only one form of humor for only one audience:  Gen Z humor Instagram captions. This type of humor tends toward absurdities, which can be easier to generate than something that needs to be logically sound. Being illogically surprising is probably easier than being logical and surprising. Future work would have to test whether similar techniques work on other humor tastes. Some of our techniques, like fine-tuning, would likely work generically for all humor types, but other skills might need to be tried.

The caption humor is difficult, but it is more well-defined than other forms of humor. Caption humor only requires a punchline for a given image (the setup). Other forms of humor like standup comedy and popular humor magazines require generating both the setup and the punchline. A future direction is to explore what additional "skills" are needed to generate jokes with both setup and punchline. 

The humor generated here is for a public audience, but most humor made spontaneously is made for friends, and often users insider knowledge about the friends, their background, and their shared experiences. LLMs would likely struggle to make in-joke humor without a source of inside information to train on.

In our baseline captions, we crafted a simple prompt for GPT-4o to write humor. It is possible that with a better prompt or multiple generations, one could generate similar results. However, that is effectively what the system performs. It might be possible that there are prompts that don't employ any humor skills that can also generate jokes funnier than baseline GPT. Future work should test more prompts - both with and without skills to see if there are approaches other than skills that can enhance LLM humor generation.


\section{Conclusion}
Humor is a highly valued human skill - with the power to teach, to entertain, and to connect. Humor is complex because generating it requires understanding human social, cultural, and emotional experiences.
In this paper, we study whether giving LLMs human-like skills for humor generation can improve its abilities to write effective humor. Since humor is often about finding relatable connections between people, we focus on generating humor for one group and type of humor: Gen Z Instagram caption humor. We run 3 studies showing that LLMs with humor skills are rated as more funny than LLMs alone, and as funny as the top-rated human-written captions. This points to a future where LLMs can potentially write with similar social skills as people to produce emotional reactions and human bonding.

\bibliographystyle{ACM-Reference-Format}
\bibliography{sample-base}


\begin{thebibliography}{39}


\ifx \showCODEN    \undefined \def \showCODEN     #1{\unskip}     \fi
\ifx \showDOI      \undefined \def \showDOI       #1{#1}\fi
\ifx \showISBNx    \undefined \def \showISBNx     #1{\unskip}     \fi
\ifx \showISBNxiii \undefined \def \showISBNxiii  #1{\unskip}     \fi
\ifx \showISSN     \undefined \def \showISSN      #1{\unskip}     \fi
\ifx \showLCCN     \undefined \def \showLCCN      #1{\unskip}     \fi
\ifx \shownote     \undefined \def \shownote      #1{#1}          \fi
\ifx \showarticletitle \undefined \def \showarticletitle #1{#1}   \fi
\ifx \showURL      \undefined \def \showURL       {\relax}        \fi
\providecommand\bibfield[2]{#2}
\providecommand\bibinfo[2]{#2}
\providecommand\natexlab[1]{#1}
\providecommand\showeprint[2][]{arXiv:#2}

\bibitem[Anonymous(2024)]%
        {anonymous2024funlms}
\bibfield{author}{\bibinfo{person}{Anonymous}.} \bibinfo{year}{2024}\natexlab{}.
\newblock \showarticletitle{Fun{LM}s: Methods for Fine-tuning {LLM}s to Generate Humor}. In \bibinfo{booktitle}{\emph{Submitted to ACL Rolling Review - June 2024}}.
\newblock
\urldef\tempurl%
\url{https://openreview.net/forum?id=JMwjgXHLCj}
\showURL{%
\tempurl}
\newblock
\shownote{under review}.


\bibitem[Binsted and Ritchie(1997)]%
        {jape}
\bibfield{author}{\bibinfo{person}{Kim Binsted} {and} \bibinfo{person}{Graeme Ritchie}.} \bibinfo{year}{1997}\natexlab{}.
\newblock \showarticletitle{Computational humor}. In \bibinfo{booktitle}{\emph{Proceedings of the 14th International Joint Conference on Artificial Intelligence (IJCAI)}}, Vol.~\bibinfo{volume}{97}. \bibinfo{pages}{1089--1094}.
\newblock


\bibitem[Brodsky(2024)]%
        {ai_self_eval}
\bibfield{author}{\bibinfo{person}{Sascha Brodsky}.} \bibinfo{year}{2024}\natexlab{}.
\newblock \bibinfo{title}{Who watches the AI watchers? The challenge of self-evaluating AI}.
\newblock \bibinfo{howpublished}{\url{https://www.ibm.com/think/news/ai-testing-advances}}.
\newblock
\newblock
\shownote{Accessed: 2025-02-06}.


\bibitem[Cai et~al\mbox{.}(2024)]%
        {cai2024surveymixtureexperts}
\bibfield{author}{\bibinfo{person}{Weilin Cai}, \bibinfo{person}{Juyong Jiang}, \bibinfo{person}{Fan Wang}, \bibinfo{person}{Jing Tang}, \bibinfo{person}{Sunghun Kim}, {and} \bibinfo{person}{Jiayi Huang}.} \bibinfo{year}{2024}\natexlab{}.
\newblock \bibinfo{title}{A Survey on Mixture of Experts}.
\newblock
\newblock
\showeprint[arxiv]{2407.06204}~[cs.LG]
\urldef\tempurl%
\url{https://arxiv.org/abs/2407.06204}
\showURL{%
\tempurl}


\bibitem[Carter(2001)]%
        {Carter2001}
\bibfield{author}{\bibinfo{person}{Judy Carter}.} \bibinfo{year}{2001}\natexlab{}.
\newblock \bibinfo{booktitle}{\emph{{The Comedy Bible: From Stand-up to Sitcom--The Comedy Writer's Ultimate "How To" Guide}}}.
\newblock \bibinfo{publisher}{Touchstone}, \bibinfo{address}{New York, New York, USA}. 368 pages.
\newblock
\showISBNx{0743201256}


\bibitem[Cerullo(2024)]%
        {ai_scams}
\bibfield{author}{\bibinfo{person}{Megan Cerullo}.} \bibinfo{year}{2024}\natexlab{}.
\newblock \bibinfo{title}{https://www.cbsnews.com/news/elder-scams-family-safe-word/}.
\newblock \bibinfo{howpublished}{\url{https://www.cbsnews.com/news/elder-scams-family-safe-word/}}.
\newblock
\newblock
\shownote{Accessed: 2025-02-06}.


\bibitem[Chen et~al\mbox{.}(2019)]%
        {smartcompose}
\bibfield{author}{\bibinfo{person}{Mia~Xu Chen}, \bibinfo{person}{Benjamin~N. Lee}, \bibinfo{person}{Gagan Bansal}, \bibinfo{person}{Yuan Cao}, \bibinfo{person}{Shuyuan Zhang}, \bibinfo{person}{Justin Lu}, \bibinfo{person}{Jackie Tsay}, \bibinfo{person}{Yinan Wang}, \bibinfo{person}{Andrew~M. Dai}, \bibinfo{person}{Zhifeng Chen}, \bibinfo{person}{Timothy Sohn}, {and} \bibinfo{person}{Yonghui Wu}.} \bibinfo{year}{2019}\natexlab{}.
\newblock \showarticletitle{Gmail Smart Compose: Real-Time Assisted Writing}. In \bibinfo{booktitle}{\emph{Proceedings of the 25th ACM SIGKDD International Conference on Knowledge Discovery \& Data Mining}} (Anchorage, AK, USA) \emph{(\bibinfo{series}{KDD '19})}. \bibinfo{publisher}{Association for Computing Machinery}, \bibinfo{address}{New York, NY, USA}, \bibinfo{pages}{2287–2295}.
\newblock
\showISBNx{9781450362016}
\urldef\tempurl%
\url{https://doi.org/10.1145/3292500.3330723}
\showDOI{\tempurl}


\bibitem[Dean(2000a)]%
        {Dean2000}
\bibfield{author}{\bibinfo{person}{Greg Dean}.} \bibinfo{year}{2000}\natexlab{a}.
\newblock \bibinfo{booktitle}{\emph{{Step by Step to Stand-Up Comedy}}}.
\newblock \bibinfo{publisher}{Heinemann Drama}, \bibinfo{address}{Portsmouth, New Hampshire}. 208 pages.
\newblock
\showISBNx{0325001790}


\bibitem[Dean(2000b)]%
        {humor_Dean2000}
\bibfield{author}{\bibinfo{person}{Greg Dean}.} \bibinfo{year}{2000}\natexlab{b}.
\newblock \bibinfo{booktitle}{\emph{{Step by Step to Stand-Up Comedy}}}.
\newblock \bibinfo{publisher}{Heinemann Drama}, \bibinfo{address}{Portsmouth, New Hampshire}. 208 pages.
\newblock
\showISBNx{0325001790}


\bibitem[DiResta and Goldstein(2024)]%
        {diresta2024spammersscammersleverageaigenerated}
\bibfield{author}{\bibinfo{person}{Renee DiResta} {and} \bibinfo{person}{Josh~A. Goldstein}.} \bibinfo{year}{2024}\natexlab{}.
\newblock \bibinfo{title}{How Spammers and Scammers Leverage AI-Generated Images on Facebook for Audience Growth}.
\newblock
\newblock
\showeprint[arxiv]{2403.12838}~[cs.CY]
\urldef\tempurl%
\url{https://arxiv.org/abs/2403.12838}
\showURL{%
\tempurl}


\bibitem[Gorenz and Schwarz(2024)]%
        {gptvsturk}
\bibfield{author}{\bibinfo{person}{Drew Gorenz} {and} \bibinfo{person}{Norbert Schwarz}.} \bibinfo{year}{2024}\natexlab{}.
\newblock \showarticletitle{How funny is ChatGPT? A comparison of human- and A.I.-produced jokes}.
\newblock \bibinfo{journal}{\emph{PLOS ONE}} \bibinfo{volume}{19}, \bibinfo{number}{7} (\bibinfo{date}{07} \bibinfo{year}{2024}), \bibinfo{pages}{1--13}.
\newblock
\urldef\tempurl%
\url{https://doi.org/10.1371/journal.pone.0305364}
\showDOI{\tempurl}


\bibitem[He et~al\mbox{.}(2019)]%
        {he2019pungenerationsurprise}
\bibfield{author}{\bibinfo{person}{He He}, \bibinfo{person}{Nanyun Peng}, {and} \bibinfo{person}{Percy Liang}.} \bibinfo{year}{2019}\natexlab{}.
\newblock \bibinfo{title}{Pun Generation with Surprise}.
\newblock
\newblock
\showeprint[arxiv]{1904.06828}~[cs.CL]
\urldef\tempurl%
\url{https://arxiv.org/abs/1904.06828}
\showURL{%
\tempurl}


\bibitem[Holloway(2010)]%
        {Holloway2010}
\bibfield{author}{\bibinfo{person}{Sally Holloway}.} \bibinfo{year}{2010}\natexlab{}.
\newblock \bibinfo{booktitle}{\emph{{The Serious Guide to Joke Writing: How To Say Something Funny About Anything}}}.
\newblock \bibinfo{publisher}{Bookshaker}, \bibinfo{address}{Great Yarmouth, UK}. 207 pages.
\newblock
\showISBNx{1907498370}


\bibitem[Horvitz et~al\mbox{.}(2024)]%
        {unfun}
\bibfield{author}{\bibinfo{person}{Zachary Horvitz}, \bibinfo{person}{Jingru Chen}, \bibinfo{person}{Rahul Aditya}, \bibinfo{person}{Harshvardhan Srivastava}, \bibinfo{person}{Robert West}, \bibinfo{person}{Zhou Yu}, {and} \bibinfo{person}{Kathleen McKeown}.} \bibinfo{year}{2024}\natexlab{}.
\newblock \bibinfo{title}{Getting Serious about Humor: Crafting Humor Datasets with Unfunny Large Language Models}.
\newblock
\newblock
\showeprint[arxiv]{2403.00794}~[cs.CL]
\urldef\tempurl%
\url{https://arxiv.org/abs/2403.00794}
\showURL{%
\tempurl}


\bibitem[Hurley et~al\mbox{.}(2011)]%
        {hurley2011inside}
\bibfield{author}{\bibinfo{person}{Matthew~M. Hurley}, \bibinfo{person}{Daniel~C. Dennett}, {and} \bibinfo{person}{Reginald~B. Adams~Jr.}} \bibinfo{year}{2011}\natexlab{}.
\newblock \bibinfo{booktitle}{\emph{Inside Jokes: Using Humor to Reverse-Engineer the Mind}}.
\newblock \bibinfo{publisher}{The MIT Press}, \bibinfo{address}{Cambridge, MA, USA}.
\newblock


\bibitem[Jakesch et~al\mbox{.}(2023)]%
        {naaman_opinion}
\bibfield{author}{\bibinfo{person}{Maurice Jakesch}, \bibinfo{person}{Advait Bhat}, \bibinfo{person}{Daniel Buschek}, \bibinfo{person}{Lior Zalmanson}, {and} \bibinfo{person}{Mor Naaman}.} \bibinfo{year}{2023}\natexlab{}.
\newblock \showarticletitle{Co-Writing with Opinionated Language Models Affects Users’ Views}. In \bibinfo{booktitle}{\emph{Proceedings of the 2023 CHI Conference on Human Factors in Computing Systems}} (Hamburg, Germany) \emph{(\bibinfo{series}{CHI '23})}. \bibinfo{publisher}{Association for Computing Machinery}, \bibinfo{address}{New York, NY, USA}, Article \bibinfo{articleno}{111}, \bibinfo{numpages}{15}~pages.
\newblock
\showISBNx{9781450394215}
\urldef\tempurl%
\url{https://doi.org/10.1145/3544548.3581196}
\showDOI{\tempurl}


\bibitem[Jentzsch and Kersting(2023)]%
        {gptnotfunny}
\bibfield{author}{\bibinfo{person}{Sophie Jentzsch} {and} \bibinfo{person}{Kristian Kersting}.} \bibinfo{year}{2023}\natexlab{}.
\newblock \bibinfo{title}{ChatGPT is fun, but it is not funny! Humor is still challenging Large Language Models}.
\newblock
\newblock
\showeprint[arxiv]{2306.04563}~[cs.AI]
\urldef\tempurl%
\url{https://arxiv.org/abs/2306.04563}
\showURL{%
\tempurl}


\bibitem[Kaplan(2013)]%
        {Kaplan2013}
\bibfield{author}{\bibinfo{person}{Steve Kaplan}.} \bibinfo{year}{2013}\natexlab{}.
\newblock \bibinfo{booktitle}{\emph{{The Hidden Tools of Comedy: The Serious Business of Being Funny}}}.
\newblock \bibinfo{publisher}{Michael Wiese Productions}, \bibinfo{address}{Studio City, CA}. 280 pages.
\newblock
\showISBNx{1615931406}


\bibitem[Kiddon and Brun(2011)]%
        {twss}
\bibfield{author}{\bibinfo{person}{Chlo{\'e} Kiddon} {and} \bibinfo{person}{Yuriy Brun}.} \bibinfo{year}{2011}\natexlab{}.
\newblock \showarticletitle{That's What She Said: Double Entendre Identification}. In \bibinfo{booktitle}{\emph{Proceedings of the 49th Annual Meeting of the Association for Computational Linguistics: Human Language Technologies: Short Papers - Volume 2}} (Portland, Oregon) \emph{(\bibinfo{series}{HLT '11})}. \bibinfo{publisher}{Association for Computational Linguistics}, \bibinfo{address}{Stroudsburg, PA, USA}, \bibinfo{pages}{89--94}.
\newblock
\showISBNx{978-1-932432-88-6}
\urldef\tempurl%
\url{http://dl.acm.org/citation.cfm?id=2002736.2002756}
\showURL{%
\tempurl}


\bibitem[Ma et~al\mbox{.}(2023)]%
        {ma2023letsthoughtexperimentusing}
\bibfield{author}{\bibinfo{person}{Xiao Ma}, \bibinfo{person}{Swaroop Mishra}, \bibinfo{person}{Ahmad Beirami}, \bibinfo{person}{Alex Beutel}, {and} \bibinfo{person}{Jilin Chen}.} \bibinfo{year}{2023}\natexlab{}.
\newblock \bibinfo{title}{Let's Do a Thought Experiment: Using Counterfactuals to Improve Moral Reasoning}.
\newblock
\newblock
\showeprint[arxiv]{2306.14308}~[cs.CL]
\urldef\tempurl%
\url{https://arxiv.org/abs/2306.14308}
\showURL{%
\tempurl}


\bibitem[Majic(2024)]%
        {majic2024rizz}
\bibfield{author}{\bibinfo{person}{Josipa Majic}.} \bibinfo{year}{2024}\natexlab{}.
\newblock \showarticletitle{Rizz App: How the 5th Most Downloaded Dating App is Redefining Digital Relationships}.
\newblock \bibinfo{journal}{\emph{Forbes}} (\bibinfo{year}{2024}).
\newblock
\urldef\tempurl%
\url{https://www.forbes.com/sites/josipamajic/2024/09/09/rizz-app-how-the-5th-most-downloaded-dating-app-is-redefining-digital-relationships/}
\showURL{%
\tempurl}
\newblock
\shownote{Accessed: 2024-09-13}.


\bibitem[Mirowski et~al\mbox{.}(2024)]%
        {deepmind_humor}
\bibfield{author}{\bibinfo{person}{Piotr~W. Mirowski}, \bibinfo{person}{Juliette Love}, \bibinfo{person}{Kory Mathewson}, {and} \bibinfo{person}{Shakir Mohamed}.} \bibinfo{year}{2024}\natexlab{}.
\newblock \showarticletitle{A Robot Walks into a Bar: Can Language Models Serve as Creativity Support Tools for Comedy? An Evaluation of LLMs’ Humour Alignment with Comedians}. In \bibinfo{booktitle}{\emph{The 2024 ACM Conference on Fairness, Accountability, and Transparency (FAccT ’24)}}. \bibinfo{publisher}{ACM}, \bibinfo{address}{Rio de Janeiro, Brazil}, \bibinfo{pages}{15}.
\newblock
\urldef\tempurl%
\url{https://doi.org/10.1145/3630106.3658993}
\showDOI{\tempurl}


\bibitem[of~Philosophy(2015)]%
        {sep_humor}
\bibfield{author}{\bibinfo{person}{Stanford~Encyclopedia of Philosophy}.} \bibinfo{year}{2015}\natexlab{}.
\newblock \bibinfo{title}{Philosophy of Humor}.
\newblock \bibinfo{howpublished}{\url{http://plato.stanford.edu/entries/humor/}}.
\newblock
\newblock
\shownote{Accessed: 2015-12-01}.


\bibitem[Park et~al\mbox{.}(2023)]%
        {joon_agents}
\bibfield{author}{\bibinfo{person}{Joon~Sung Park}, \bibinfo{person}{Joseph~C. O'Brien}, \bibinfo{person}{Carrie~J. Cai}, \bibinfo{person}{Meredith~Ringel Morris}, \bibinfo{person}{Percy Liang}, {and} \bibinfo{person}{Michael~S. Bernstein}.} \bibinfo{year}{2023}\natexlab{}.
\newblock \bibinfo{title}{Generative Agents: Interactive Simulacra of Human Behavior}.
\newblock
\newblock
\showeprint[arxiv]{2304.03442}~[cs.HC]
\urldef\tempurl%
\url{https://arxiv.org/abs/2304.03442}
\showURL{%
\tempurl}


\bibitem[Radford et~al\mbox{.}(2019)]%
        {gpt2}
\bibfield{author}{\bibinfo{person}{Alec Radford}, \bibinfo{person}{Jeffrey Wu}, \bibinfo{person}{Rewon Child}, \bibinfo{person}{David Luan}, \bibinfo{person}{Dario Amodei}, {and} \bibinfo{person}{Ilya Sutskever}.} \bibinfo{year}{2019}\natexlab{}.
\newblock \showarticletitle{Language Models are Unsupervised Multitask Learners}.
\newblock \bibinfo{journal}{\emph{OpenAI Blog}} (\bibinfo{year}{2019}).
\newblock
\urldef\tempurl%
\url{https://cdn.openai.com/better-language-models/language_models_are_unsupervised_multitask_learners.pdf}
\showURL{%
\tempurl}


\bibitem[Raskin(2009a)]%
        {humor_Raskin2009}
\bibfield{author}{\bibinfo{person}{Victor Raskin}.} \bibinfo{year}{2009}\natexlab{a}.
\newblock \bibinfo{booktitle}{\emph{{The Primer of Humor Research}}}.
\newblock \bibinfo{publisher}{De Gruyter}, \bibinfo{address}{Berlin, Germany}. 673 pages.
\newblock
\showISBNx{3110198495}


\bibitem[Raskin(2009b)]%
        {Raskin2009}
\bibfield{author}{\bibinfo{person}{Victor Raskin}.} \bibinfo{year}{2009}\natexlab{b}.
\newblock \bibinfo{booktitle}{\emph{{The Primer of Humor Research}}}.
\newblock \bibinfo{publisher}{De Gruyter}, \bibinfo{address}{Berlin, Germany}. 673 pages.
\newblock
\showISBNx{3110198495}


\bibitem[Shahaf et~al\mbox{.}(2015)]%
        {shahafjokes}
\bibfield{author}{\bibinfo{person}{Dafna Shahaf}, \bibinfo{person}{Eric Horvitz}, {and} \bibinfo{person}{Robert Mankoff}.} \bibinfo{year}{2015}\natexlab{}.
\newblock \showarticletitle{Inside Jokes: Identifying Humorous Cartoon Captions}. In \bibinfo{booktitle}{\emph{Proceedings of the 21th ACM SIGKDD International Conference on Knowledge Discovery and Data Mining}} (Sydney, NSW, Australia) \emph{(\bibinfo{series}{KDD '15})}. \bibinfo{publisher}{ACM}, \bibinfo{address}{New York, NY, USA}, \bibinfo{pages}{1065--1074}.
\newblock
\showISBNx{978-1-4503-3664-2}
\urldef\tempurl%
\url{https://doi.org/10.1145/2783258.2783388}
\showDOI{\tempurl}


\bibitem[Shinn et~al\mbox{.}(2023)]%
        {reflection}
\bibfield{author}{\bibinfo{person}{Noah Shinn}, \bibinfo{person}{Federico Cassano}, \bibinfo{person}{Ashwin Gopinath}, \bibinfo{person}{Karthik Narasimhan}, {and} \bibinfo{person}{Shunyu Yao}.} \bibinfo{year}{2023}\natexlab{}.
\newblock \showarticletitle{Reflexion: Language Agents with Verbal Reinforcement Learning}.
\newblock \bibinfo{journal}{\emph{Advances in Neural Information Processing Systems}}  \bibinfo{volume}{36} (\bibinfo{year}{2023}).
\newblock
\showISSN{1049-5258}
\newblock
\shownote{Publisher Copyright: {\textcopyright} 2023 Neural information processing systems foundation. All rights reserved.; 37th Conference on Neural Information Processing Systems, NeurIPS 2023 ; Conference date: 10-12-2023 Through 16-12-2023}.


\bibitem[Suh et~al\mbox{.}(2021)]%
        {socialglue}
\bibfield{author}{\bibinfo{person}{Minhyang~(Mia) Suh}, \bibinfo{person}{Emily Youngblom}, \bibinfo{person}{Michael Terry}, {and} \bibinfo{person}{Carrie~J Cai}.} \bibinfo{year}{2021}\natexlab{}.
\newblock \showarticletitle{AI as Social Glue: Uncovering the Roles of Deep Generative AI during Social Music Composition}. In \bibinfo{booktitle}{\emph{Proceedings of the 2021 CHI Conference on Human Factors in Computing Systems}} (Yokohama, Japan) \emph{(\bibinfo{series}{CHI '21})}. \bibinfo{publisher}{Association for Computing Machinery}, \bibinfo{address}{New York, NY, USA}, Article \bibinfo{articleno}{582}, \bibinfo{numpages}{11}~pages.
\newblock
\showISBNx{9781450380966}
\urldef\tempurl%
\url{https://doi.org/10.1145/3411764.3445219}
\showDOI{\tempurl}


\bibitem[Taylor and Mazlack(2004)]%
        {Taylor04computationallyrecognizing}
\bibfield{author}{\bibinfo{person}{Julia~M. Taylor} {and} \bibinfo{person}{Lawrence~J. Mazlack}.} \bibinfo{year}{2004}\natexlab{}.
\newblock \showarticletitle{Computationally recognizing wordplay in jokes}. In \bibinfo{booktitle}{\emph{In Proceedings of CogSci 2004}}.
\newblock


\bibitem[Tikhonov and Shtykovskiy(2024)]%
        {tikhonov2024humormechanicsadvancinghumor}
\bibfield{author}{\bibinfo{person}{Alexey Tikhonov} {and} \bibinfo{person}{Pavel Shtykovskiy}.} \bibinfo{year}{2024}\natexlab{}.
\newblock \bibinfo{title}{Humor Mechanics: Advancing Humor Generation with Multistep Reasoning}.
\newblock
\newblock
\showeprint[arxiv]{2405.07280}~[cs.CL]
\urldef\tempurl%
\url{https://arxiv.org/abs/2405.07280}
\showURL{%
\tempurl}


\bibitem[Toplyn(2023a)]%
        {witscript2}
\bibfield{author}{\bibinfo{person}{Joe Toplyn}.} \bibinfo{year}{2023}\natexlab{a}.
\newblock \bibinfo{title}{Witscript 2: A System for Generating Improvised Jokes Without Wordplay}.
\newblock
\newblock
\showeprint[arxiv]{2302.03036}~[cs.CL]
\urldef\tempurl%
\url{https://arxiv.org/abs/2302.03036}
\showURL{%
\tempurl}


\bibitem[Toplyn(2023b)]%
        {witscript3}
\bibfield{author}{\bibinfo{person}{Joe Toplyn}.} \bibinfo{year}{2023}\natexlab{b}.
\newblock \bibinfo{title}{Witscript 3: A Hybrid AI System for Improvising Jokes in a Conversation}.
\newblock
\newblock
\showeprint[arxiv]{2301.02695}~[cs.CL]
\urldef\tempurl%
\url{https://arxiv.org/abs/2301.02695}
\showURL{%
\tempurl}


\bibitem[Toplyn(2023c)]%
        {witscript1}
\bibfield{author}{\bibinfo{person}{Joe Toplyn}.} \bibinfo{year}{2023}\natexlab{c}.
\newblock \bibinfo{title}{Witscript: A System for Generating Improvised Jokes in a Conversation}.
\newblock
\newblock
\showeprint[arxiv]{2302.02008}~[cs.CL]
\urldef\tempurl%
\url{https://arxiv.org/abs/2302.02008}
\showURL{%
\tempurl}


\bibitem[Vorhaus(1994)]%
        {Vorhaus1994}
\bibfield{author}{\bibinfo{person}{John Vorhaus}.} \bibinfo{year}{1994}\natexlab{}.
\newblock \bibinfo{booktitle}{\emph{{The Comic Toolbox: How to Be Funny Even If You're Not}}}.
\newblock \bibinfo{publisher}{Silman James Press}, \bibinfo{address}{Los Angeles, CA}. 191 pages.
\newblock
\showISBNx{1879505215}


\bibitem[Wei et~al\mbox{.}(2024)]%
        {cot}
\bibfield{author}{\bibinfo{person}{Jason Wei}, \bibinfo{person}{Xuezhi Wang}, \bibinfo{person}{Dale Schuurmans}, \bibinfo{person}{Maarten Bosma}, \bibinfo{person}{Brian Ichter}, \bibinfo{person}{Fei Xia}, \bibinfo{person}{Ed~H. Chi}, \bibinfo{person}{Quoc~V. Le}, {and} \bibinfo{person}{Denny Zhou}.} \bibinfo{year}{2024}\natexlab{}.
\newblock \showarticletitle{Chain-of-thought prompting elicits reasoning in large language models}. In \bibinfo{booktitle}{\emph{Proceedings of the 36th International Conference on Neural Information Processing Systems}} (New Orleans, LA, USA) \emph{(\bibinfo{series}{NIPS '22})}. \bibinfo{publisher}{Curran Associates Inc.}, \bibinfo{address}{Red Hook, NY, USA}, Article \bibinfo{articleno}{1800}, \bibinfo{numpages}{14}~pages.
\newblock
\showISBNx{9781713871088}


\bibitem[Wu et~al\mbox{.}(2022)]%
        {cai_ai_chains}
\bibfield{author}{\bibinfo{person}{Tongshuang Wu}, \bibinfo{person}{Michael Terry}, {and} \bibinfo{person}{Carrie~Jun Cai}.} \bibinfo{year}{2022}\natexlab{}.
\newblock \showarticletitle{AI Chains: Transparent and Controllable Human-AI Interaction by Chaining Large Language Model Prompts}. In \bibinfo{booktitle}{\emph{Proceedings of the 2022 CHI Conference on Human Factors in Computing Systems}} (New Orleans, LA, USA) \emph{(\bibinfo{series}{CHI '22})}. \bibinfo{publisher}{Association for Computing Machinery}, \bibinfo{address}{New York, NY, USA}, Article \bibinfo{articleno}{385}, \bibinfo{numpages}{22}~pages.
\newblock
\showISBNx{9781450391573}
\urldef\tempurl%
\url{https://doi.org/10.1145/3491102.3517582}
\showDOI{\tempurl}


\bibitem[Young et~al\mbox{.}(2014)]%
        {young2014image}
\bibfield{author}{\bibinfo{person}{Peter Young}, \bibinfo{person}{Alice Lai}, \bibinfo{person}{Micah Hodosh}, {and} \bibinfo{person}{Julia Hockenmaier}.} \bibinfo{year}{2014}\natexlab{}.
\newblock \showarticletitle{From image descriptions to visual denotations: New similarity metrics for semantic inference over event descriptions}. In \bibinfo{booktitle}{\emph{Transactions of the Association for Computational Linguistics}}, Vol.~\bibinfo{volume}{2}. \bibinfo{pages}{67--78}.
\newblock


\end{thebibliography}

\end{document}